# Article Processing Charges, Altmetrics and Citation Impact: Is there an economic rationale?

Abdelghani Maddi[1], David Sapinho[2]

*[1] abdelghani.maddi@hceres.fr*
Observatoire des Sciences et Techniques, Hcéres, 2 Rue Albert Einstein, Paris, 75013 France.

*[2] david.sapinho@hceres.fr*
Observatoire des Sciences et Techniques, Hcéres, 2 Rue Albert Einstein, Paris, 75013 France

## Abstract
The present study aims to analyze 1) the relationship between Citation Normalized Score of scientific publications and Article Processing Charges (APCs) of Gold Open Access (OA) publications 2) the determinants of APCs. To do so, we used APCs information provided by the OpenAPC database, citation scores of publications from the WoS database and, for Altmetrics, data from Altmetrics.com database, over the period from 2006 to 2019 for 83,752 articles published in 4751 journals belonging to 267 distinct publishers. Results show that contrary to common belief, paying high APCs does not necessarily increase the impact of publications. First, large publishers with high impact are not the most expensive. Second, publishers with the highest APCs are not necessarily the best in terms of impact. Correlation between APCs and impact is moderate. Regarding the determinants, results indicate that APCs are on average 50% higher in hybrid journals than in full OA journals. The results also suggest that Altmetrics do not have a great impact: OA articles that have garnered the most attention on internet are articles with relatively low APCs. Another interesting result is that the "number of readers" indicator is more effective as it is more correlated with classic bibliometrics indicators than the Altmetrics score.

## Keywords
Article Processing Charges; Citation impact; Altmetrics; publisher; Open Access; bibliometric indicators.

## Acknowledgement
The present paper is a substantially extended version of the contribution (Maddi and Sapinho, 2021) presented at the ISSI 2021 – 18th International Conference on Scientometrics & Informetrics – Leuven, Belgium.
The authors would like to thank the anonymous referees and the guest-editor for their very helpful comments, which have significantly improved the quality of this paper.

---

[1] Corresponding author : Abdelghani Maddi, Observatoire des Sciences et Techniques, Hcéres, 2 Rue Albert Einstein, Paris, 75013, France, T. 33 (0)1 55 55 61 48.

Introduction

Since the start of the 21st century, the scientific community has witnessed an unprecedented popularity of the Open Access (OA) movement (Björk, 2004; Chen-Chi, 2006; Sotudeh and Estakhr, 2018). A successful transition into an OA publishing model is seen as a good way to ensure better dissemination of knowledge and more equity between researchers, especially for those from institutions in low-income countries faced with the issue of paying subscription fees (Prosser, 2003; Tananbaum, 2003; Solomon and Björk, 2012; Cary and Rockwell, 2020).

However, OA does not necessarily mean "free", and raises the question of the business model underlying scientific publishing. For institutions, it may even generate new costs: in addition to the subscription fees, they are increasingly led to pay costs of OA publication (Maddi, 2020; Maddi and Sapinho, 2021). This concerns a part of Gold OA publications which are based on the "author-pays" business model (Rizor and Holley, 2014; Sotudeh, Ghasempour and Yaghtin, 2015). Thus, authors pay the "Article Processing Charges" (APCs), usually via their institution or via their project money, to allow open access to the publication (Asai, 2019; Khoo, 2019; Bruns, Rimmert and Taubert, 2020; Copiello, 2020).

The APCs have increased significantly over time. This rise has been estimated at three times faster than it be if indexed to inflation (Khoo 2019). The trend appears to be stronger in more frequently cited journals, as highlighted for Biomed Central journals (Asai 2020), medical and specific OA journals (Asai 2019). These findings would also suggest that large subscription journal publishers tend to set higher APCs. Nevertheless, there is no evidence to date that the introduction of APCs for a given journal reduced its publications volume (Khoo, 2019). In other words, once able to pay an APC, authors give little emphasis to their amount. The APC-based publishing model is being more and more integrated by academic institutions that aspire to switch to an economic model excluding subscription fees. Thus, APCs are a considerable burden on "total cost of publication" for institutions, reaching 10% in 2013 (Pinfield, Salter and Bath, 2016).

While many studies have described relationships between OA publications and citation level (Gumpenberger, Ovalle-Perandones and Gorraiz, 2013; Zhang and Watson, 2017; Piwowar *et al.*, 2018), only few have focused on the amount of APCs, with heterogeneous findings. On one hand, APCs based journals have been regarded, in general, as more cited than other OA journals (Björk and Solomon, 2012) , on the other hand, it was concluded that both categories have, on average, similar performances with some disciplinary differences (Ghane, Niazmand and Sabet Sarvestani, 2020). Another study analyzed the relationship between APCs and scientific impact of publications using respectively DOAJ and Scopus data (Björk and Solomon, 2015). On a set of 61,081 publications and 595 journals, authors showed that there is a moderate correlation (0.4) between the two indicators at the journal level (APCs and impact). Correlation is greater (0.6) when data is weighted by the volume of articles for each journal (article level), suggesting that publishers take quality into account when pricing their journals. Likewise, authors are also sensitive to journal quality in their submission choices.

The objective of this study is twofold. First, it aims to analyze the relationship between citations normalized score of scientific publications and APCs amounts. Second, to determine whether the type of journal as well as its academic and societal impact can play a role on the determination of APCs. In other words, do the journals consider their academic and societal impact when billing APCs?

This research could be undertaken by using and crossing various sources of information using DOI. APCs for 109,141 publications, covering the period 2006 to 2019 were extracted from the OpenAPC database (https://treemaps.intact-project.org/), citations scores from our in-house Web of Science database (WoS) and Altmetrics from the Altmetric.com database. This paper is an improved version of the contribution (Maddi and Sapinho, 2021) presented at the From the initial dataset, 83,752 publications were used to study the link between APCs and impact, while 62,532 publications were used to analyze the determinants of APCs. To the extent that large, high impact publishers/journals may request high APCs, it is expected that quality will be strongly correlated with the APCs. The latter would therefore explain the publications visibility. Altmetrics scores were also expected to be correlated with APCs.

## Data

### APCs data

APCs data has been extracted from the OpenAPC database. Maintaining this database is an initiative that involves 231 institutions worldwide (5 from North America, 255 from Europe and 1 from East Asia) that publish data sets on fees paid for OA journal articles under an open database license. At the beginning of March 2020, the database contained 109,141 publications and 6,941 journals.

Each publication in this database is only assigned to the institution that implemented it, ignoring, therefore, other collaborating institutions. For more details about Open APC database see: https://treemaps.intact-project.org/page/about.html

### OST data on citation impact and disciplines

The data about citations scores and disciplinary assignation of publications has been extracted from the french Observatoire des Sciences et Techniques' (OST) in-house database. It includes five indexes of WoS available from Clarivate Analytics (SCIE, SSCI, AHCI, CPCI-SSH and CPCI-S) and corresponds to WoS content indexed through the end of March 2019. See https://clarivate.com/webofsciencegroup/solutions/webofscience-platform/. Among the 109,141 publications indexed in the OpenAPC database, 83,752 were also found in the WoS database, using the DOI.

### Altmetrics data

The data were extracted from the Altmetric.com database (https://www.altmetric.com) using the R package "rAltmetric". This package has a bug which stops the request when it encounters an unindexed DOI. The script proposed by Gorka Navarrete (https://github.com/gorkang) fixed the problem.

Of the 83,752 publications, 62,532 match the Altmetric.com database, for which all the information available on alternative metrics has been retrieved (see: https://help.altmetric.com/support/home).

### Final databases

Figure 1 schematizes the overlap between the three databases.

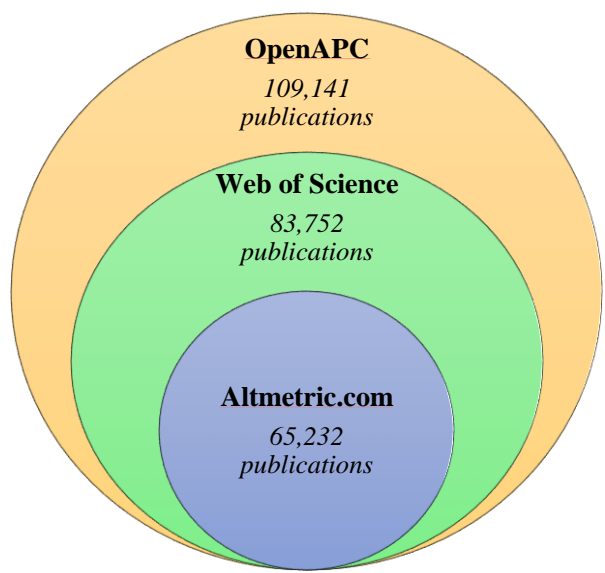

**Figure 1: Overlap between the OpenAPC, WoS and Altmetric.com databases**

*The final Web of Science dataset*

The database used for analysis includes several pieces of information about publications from the two data sources:

- From the APC data : institution that declares the publication, APCs amounts, journal in which they are published, country of the journal, publisher of journal and a flag indicating whether the journal is hybrid. By matching the OpenAPC database to that of OST.

- From the WoS-OST in-house data : we estimate the impact of publications by calculating the following indicators :

    o Normalized Citations Score (NCS) at article level: we calculated the NCS of a given article by dividing the number of citations received by the average number of citations in the same disciplines and the same year (Waltman *et al.*, 2011).

    o Mean Normalized Citation Score (MNCS) at the publisher level: we calculated the weighted average of the NCS scores, at the publisher level, based on all the articles of journals that it publishes. In the case of the OpenAPC database, the selection was restricted to OA articles only.

    o Mean Normalized Impact of Journals (MNIJ) at journal and publisher level: For each journal, we calculated the average number of citations per article in this journal for a given year, normalized by the average number of citations of all journals in the same category and year. The overall MNIJ per publisher is the average of the MNIJs by journal weighted by their respective number of publications. The MNIJ finally used is based on the citations of publications over a three-year period. MNIJ at journal level is quite different from that of Clarivate – Journal Citation Indicator (https://clarivate.com/wp-content/uploads/dlm_uploads/2021/05/Journal-Citation-Indicator-discussion-paper-2.pdf), insofar as it is calculated year by year (whereas JCI is calculated for papers published in the prior three-year period) on the one hand, and does not include the documents types in normalization on the other hand. This factor does not appear to be essential in our analysis, as the majority of the documents are articles.

- Finally, we also calculated the number of countries per article, as a measure of international collaboration.

*The final Altmetric.com dataset*

In addition to the variables presented above (from the WoS dataset), this subset of 65,232 publications contains information on alternative metrics and associated scores. For more information on the extracted variables, see the description provided here: https://cran.r-project.org/web/packages/rAltmetric/README.html.

## Method

In this article, we have carried out two models: the first one aims to analyze the link between normalized citations and APCs, while the second one analyzes the determinants of APCs including Altmetrics.

**Model 1: Citation impact and APCs**

*Dependent variable and model choice*

The dependent variable is the logarithm of Normalized Citations Score (labelled Log (NCS)) received by each publication during the period 2006-2019. To retain the zeros, we have added 1 to the NCS before making the logarithmic transformation. Log (NCS) is a continuous variable with a lower boundary at zero and an upper boundary at infinity. Thus, a left censored Tobit regression model is used to account for the disproportionate number of observations with zero values, because, indeed, a significant proportion of the observations in our sample are zeros. Tobit regressions avoid inconsistent estimates from OLS regression.

*Explanatory variable*

In this study, we seek to analyze to what extent the amount of APCs have an incidence on the number of citations received by OA scientific publications. Our explanatory variable is therefore the amount of APCs by publication.

*Control variables*

Journal impact and number of countries per publication are added as control variables. This choice was driven by the literature that shows that citations depend on journal quality and international collaboration (Maddi and Gingras, 2021). The hybrid status of a journal was also included, using a dummy variable.

**Model 2: Determinants of APCs**

*Dependent variable and model choice*

This model was built with data that integrates information on Altmetrics (the dataset of 65,232 publications published in 4099 journals). In this model, the dependent variable is the APCs by journal. Our explorations have shown that APCs by journal tends to follow a lognormal distribution. We, therefore, used a generalized linear model with a logarithmic transformation of the APCs.

*Explanatory variables*

The explanatory variables of the APCs used in the model are, respectively,

- The type of the journal (dummy variable that designates whether the journal is hybrid or fully in OA),
- The academic impact of the journal (designated by the normalized citations score)
- And finally the societal impact of the journal. Three variables are used to measure the latter: the total readers count, context journal count and Altmetrics attention score. While the first two measures provide information on the interest aroused by the magazine on the internet, the Altmetrics score indicates in a way the media *buzz* it generates (see table 1). The weights used to build this score are available in the documentation (see: https://help.altmetric.com/support/solutions/articles/6000233311-how-is-the-altmetric-attention-score-calculated-).

*Control variables*

The number of publications by discipline and by journal constitute the control variables of the model. The OST disciplinary classification (of the 254 WoS subject categories) into 11 disciplines was used for this purpose (the classification scheme is available at: https://figshare.com/articles/dataset/OST_disciplinary_classification/11897601).

## Results

In this section, we present the main results. First, we characterize the APCs data, namely: evolution of the average amount paid by institutions, characteristics of the top 20 producing publishers and then those of the most expensive one. Secondly, we present the correlation tests results between the amount of APCs on the one hand and the MNCS and MNIJ indicators on the other hand. Finally, we present the regression results.

**Overview of APCs data**

Figure 2 shows the evolution of APCs average by publication in all the OpenAPC database, and using a constant data set of publications from journals of 2006-09 period (79 journals). The interest of this approach is to distinguish the differences in pricing practices between old indexed journals and new indexed journals. Thus, several subscription-based journals began to publish OA articles from 2010 by adopting a Hybrid model. This movement has accelerated from 2013 (Besancenot and Vranceanu, 2017).

As we can see, the average amount has doubled between 2006 and 2019, going from 1,000 euros to almost 2,300 euros per publication. This is explained in particular by the new indexed journals that have been indexed into the OpenAPC database after 2009, which significantly increases the APCs average. Overall, APCs have increased significantly even for old journals (2006-09) from 1,000 to 1,800 euros on average.

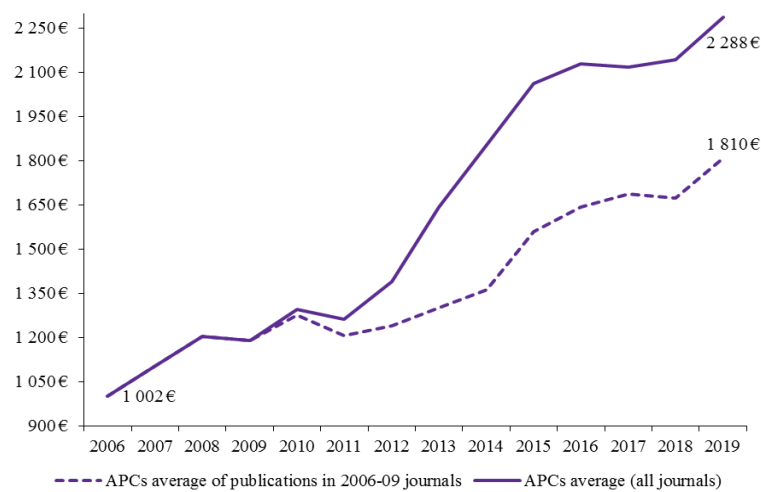

**Figure 2: APCs Average per publication, total and with constant journal set**

With all reservations that we can make on the Open APC database, we can hypothesize that this increase is notably due to a rise in demand for OA publications. As is well known in economics, increase in demand systematically leads to an increase in prices. Publishers are therefore taking advantage of this enthusiasm for OA to increase their prices.

Figure 3 describes the distribution of publications in the OpenAPC database, in relation to the amount of APCs, by discipline.

First, it highlights contrasting gold open access publication practices. This finding is similar to that observed from the WoS data base for OA publications in previous studies (Maddi, 2020; Maddi, Lardreau and Sapinho, 2021). However, the distribution is quite different than that of all the publications where the weight of engineering or chemistry, for example, is higher (OST, 2019). Thus, nearly 50% of the publications in the OpenAPC database are in Fundamental biology and Medical research, with respectively 23,466 and 19,669 publications. These two disciplines account for only 30% of publications in the WoS database (and 42% of Gold open access). The least present disciplines are mathematics, computer science and humanities. This result is largely explained by the overall size of these disciplines (see OST, 2019).

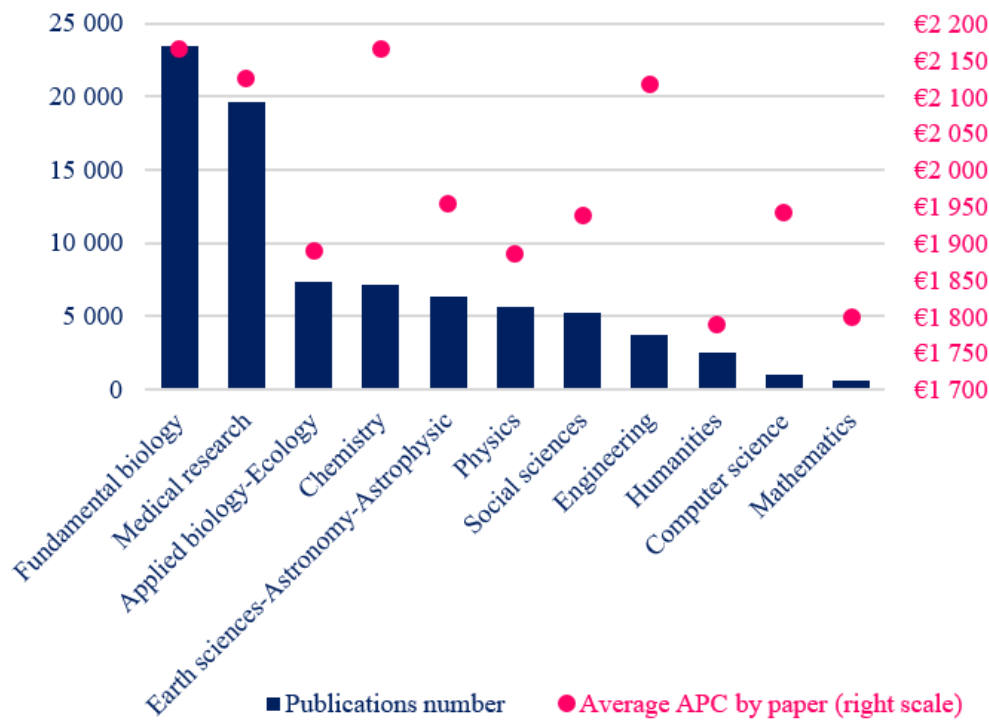

**Figure 3: publications number and average APC by discipline**

The average amount of APCs by discipline varies from € 1,800 (Mathematics and Humanities) to € 2,150 (Fundamental biology and Chemistry). The difference is therefore not significantly high depending on the discipline.

**Table 1: top 20 producing publishers: publications numbers, APCs, MNIJ and MNCS**

| Publisher | # publications | Average APCs | MNIJ | MNCS |
|---|---|---|---|---|
| Springer Nature | 14103 | 1 992 € | 1,75 | 1,29 |
| Elsevier BV | 12534 | 2 855 € | 1,99 | 1,99 |
| Public Library of Science (PLoS) | 9027 | 1 448 € | 1,46 | 1,04 |
| Wiley-Blackwell | 6959 | 2 351 € | 1,78 | 1,63 |
| Frontiers Media SA | 5725 | 1 686 € | 1,25 | 0,95 |
| MDPI AG | 3438 | 1 212 € | 1,16 | 0,85 |
| Springer Science + Business Media | 3313 | 1 536 € | 1,45 | 1,23 |
| Oxford University Press (OUP) | 3022 | 2 411 € | 2,33 | 1,97 |
| American Chemical Society (ACS) | 2299 | 2 627 € | 3,48 | 1,81 |
| IOP Publishing | 2127 | 1 569 € | 1,55 | 1,37 |
| Copernicus GmbH | 1994 | 1 492 € | 1,96 | 1,38 |
| Informa UK Limited | 1911 | 1 390 € | 0,87 | 1,47 |
| BMJ | 1604 | 2 089 € | 0,97 | 1,76 |
| Royal Society of Chemistry (RSC) | 1131 | 1 629 € | 2,49 | 1,22 |
| Optical Society of America (OSA) | 905 | 1 891 € | 1,72 | 1,91 |
| SAGE Publications | 829 | 929 € | 0,85 | 1,55 |
| Institute of Electrical & Electronics Engineers (IEEE) | 803 | 1 505 € | 1,82 | 3,12 |
| The Royal Society | 774 | 1 926 € | 1,80 | 1,48 |
| Hindawi Publishing Corporation | 739 | 1 370 € | 0,67 | 0,53 |
| Ovid Technologies (Wolters Kluwer Health) | 715 | 3 215 € | 1,41 | 1,82 |

Table 1 shows that more than half of publications are concentrated in the top 5 publishers. With a few exceptions, APCs are lower for large publishers than for the overall average. The MNIJ is higher than the world average for almost all publishers. Likewise for the MNCS. Furthermore, Table 1 shows that for some publishers, impact of publications (MNCS) is much higher than that of journals (MNIJ). This is particularly the case for the publishers "Informa UK Limited" and "BMJ". Explanation can be found in the fact that the majority of journals for these publishers (respectively, 87 and 93%) are either closed or hybrid. As demonstrated in the literature, OA publications are more cited than non-OA ones (Eysenbach, 2006; Bautista-Puig *et al.*, 2020; Ghane, Niazmand and Sabet Sarvestani, 2020). Consequently, MNCS of publications indexed in the OpenAPC database would be systematically higher than the average impact of journals in which they are published.

Table 2 shows that among the largest publishers, only Elsevier BV is listed in the top 20 most expensive ones (20th position), that are mostly American. We can also note that both impact of publications and impact of journals are high. Some exceptions can be made, especially for "MyJove Corporation" publisher with an average amount of 3081 euros for a very low impact. Similarly, "The American Association of Immunologists" charges for expensive APCs, while the average impact of journals and publications is at the level of the world average.

**Table 2: top 20 most expensive publishers: publications numbers, APCs, MNIJ and MNCS**

| Publisher | # publications | Average APCs | MNIJ | MNCS |
|---|---|---|---|---|
| American Society for Nutrition | 47 | 4 761 € | 2,34 | 2,32 |
| American Medical Association (AMA) | 28 | 4 624 € | 2,70 | 5,73 |
| American Society of Clinical Oncology (ASCO) | 23 | 4 588 € | 1,35 | 2,63 |
| Rockefeller University Press | 66 | 4 466 € | 3,64 | 2,32 |
| American Psychological Association (APA) | 132 | 3 754 € | 1,45 | 1,86 |
| American Society for Clinical Investigation | 96 | 3 656 € | 3,49 | 2,80 |
| Royal College of Psychiatrists | 64 | 3 630 € | 1,45 | 1,98 |
| EMBO | 162 | 3 410 € | 3,13 | 2,29 |
| European Respiratory Society (ERS) | 29 | 3 293 € | 1,44 | 1,80 |
| Ovid Technologies (Wolters Kluwer Health) | 715 | 3 215 € | 1,41 | 1,82 |
| American Association for Cancer Research (AACR) | 73 | 3 175 € | 2,05 | 1,45 |
| Nature Publishing Group | 350 | 3 115 € | 3,97 | 2,31 |
| American Association for the Advancement of Science (AAAS) | 166 | 3 085 € | 4,97 | 3,36 |
| MyJove Corporation | 110 | 3 081 € | 0,33 | 0,26 |
| The Company of Biologists | 414 | 3 017 € | 1,76 | 1,08 |
| The Endocrine Society | 135 | 3 017 € | 2,09 | 1,42 |
| Society for Neuroscience | 233 | 2 982 € | 2,53 | 1,56 |
| The American Association of Immunologists | 87 | 2 976 € | 1,08 | 1,06 |
| Mary Ann Liebert Inc | 117 | 2 872 € | 1,21 | 1,08 |
| Elsevier BV | 12534 | 2 855 € | 1,99 | 1,99 |

**APCs and Citation impact**

*Correlation test*

We performed a Spearman correlation test on four variables at publisher level: publications number (pub_nbr), APCs average (APCs_avg), MNIJ and MNCS. The results are presented in figure 4.

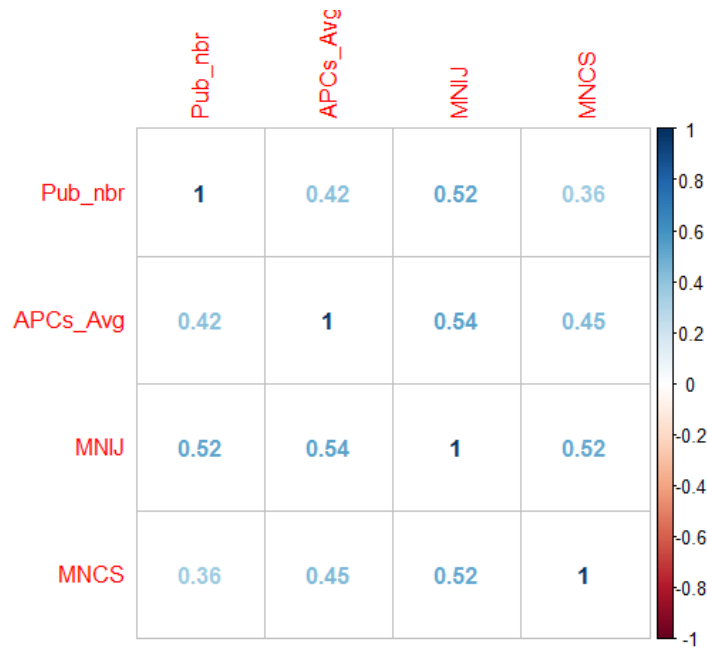

**Figure 4: Spearman correlation matrix at publisher level**

All coefficients are significant at 5%. We observe that the number of publications is more correlated with the MNIJ (0.52) than with the MNCS (0.36). This means that within the OpenAPC database there are small publishers whose publications have high citations scores and large publishers with low citations scores. We also note that the number of publications per publisher is moderately correlated with the amount of APCs. This would mean that there is no evidence of relationship between the size of publisher and the amount of APCs.

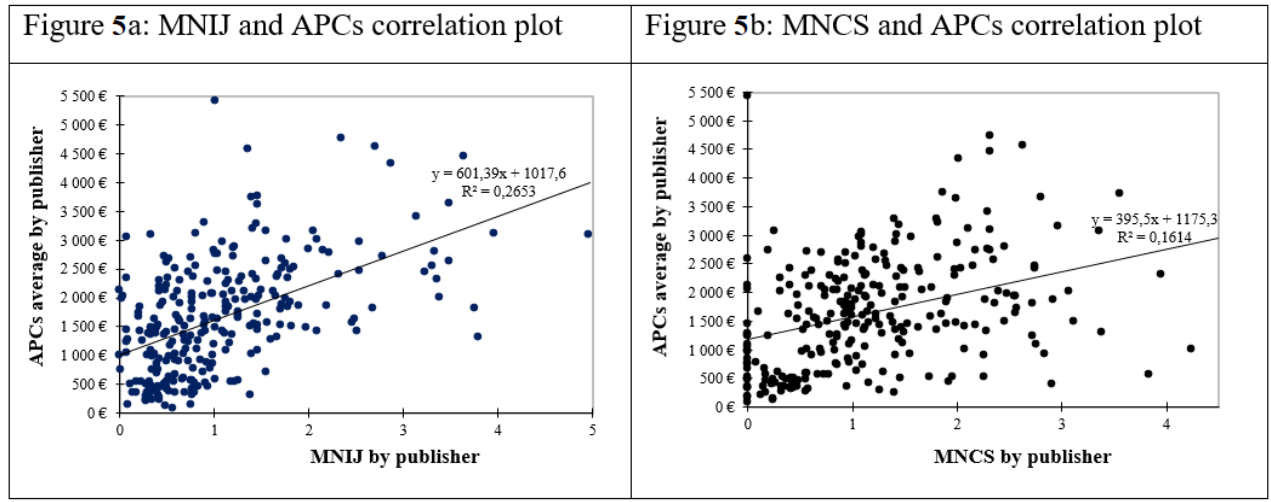

**Figure 5: APCs average per publication**

Figure 5 shows correlation plots between on the one hand MNIJ and APCs (figure 5a), and on the other hand the MNCS and the APCs (figure 5b). The correlation between MNIJ and the amount of APCs is much higher (0.54 against 0.45). This shows that publishers take the quality into account when pricing their journals. However, prices do not necessarily translate into impact, as long as the APCs are only moderately correlated with the MNCS.

*Regression results*

Table 3 summarizes Tobit regression results for explaining NCS by the amount of APCs. Regression was carried out in two stages. First, only the explanatory variable was integrated (*Log APC* per publication - model 1). Then, control variables were added (model 2).

**Table 3: Tobit maximum likelihood estimation for log NCS**

| Variables_type | Variables | Model_1 | | Model_2 | |
|---|---|---|---|---|---|
| | | Coefficient | Pr(>\|z\|) | Coefficient | Pr(>\|z\|) |
| Explanatory | $Log\ (APCs)$ | 0.22*** | $<2.22e^{-16}$ | 0.012*** | $5.25e^{-05}$ |
| Control | $Log(journal\_impact)$ | - | - | 0.471*** | $<2e^{-16}$ |
| | $Log(countries\_nbr)$ | - | - | 0.100*** | $<2e^{-16}$ |
| | $Is\_hybrid$ | - | - | 0.204*** | $<2e^{-16}$ |
| Model statistics | Wald-statistic | 2667 | $<2.22e^{-16}$ | $1.238e^{+04}$ | $<2.22e^{-16}$ |
| | Log-likelihood | $-8.063e^{+04}$ | $<2.22e^{-16}$ | $-7.607e^{+04}$ | $<2.22e^{-16}$ |
| | #publications | 83,753 | | | |
| | #Left-censored | 11907 | | | |
| | #Uncensored | 71846 | | | |
| | #Right-censored | 0 | | | |

*\*\*\*Significant at 1%*

Table 3 shows that when control variables are not taken into account, the APCs strongly affects citation score (model 1). Once control variables are integrated, the effect of the APCs drops significantly but still has a positive and significant impact on citations. Another interesting result is the impact of hybrid journals. Thus, if the journal is hybrid, citations score is higher. In other words, OA articles published in hybrid journals are generally more cited than OA articles published in 100% APCs journals. This is to be expected, given that not all well-established journals in the market have adopted a fully OA business model (Traag and Waltman, 2019). On the other hand, the main large publishers have massively integrated the hybrid model from 2013 (Besancenot and Vranceanu, 2017). In contrast, many fully OA journals are recently created journals that still have not built such a strong reputation for quality (some might even be aiming for average-level reputation and impact if this maximizes income – see (Traag and Waltman, 2019)). Hybrid journals are therefore more likely to be, at moment, in a virtuous circle where they receive higher quality manuscripts than fully OA journals, which translates to higher NCS of the published articles.

**The determinants of APCs**

In this section, the corpus including data on Altmetrics has been used (the dataset of 65,232 publications published in 4099 journals). This section consists of three parts: 1) descriptive statistics on the distribution by discipline of the average Altmetrics scores per article, the average number of readers per article and the average APCs in this dataset. 2) Correlation tests, 3) and finally the results of the multiple linear regression on the determinants of APCs.

*Descriptive statistics*

**Table 4: APCs, Altmetrics score and readers count by discipline (rounded values)**

| Discipline | Altmetrics score (average per article) | Readers count (average per article) | APCs (average per article) |
|---|---|---|---|
| Multidisciplinary | 34 | 71 | 1 680 € |
| Social Sciences | 31 | 99 | 1 977 € |
| Humanities | 30 | 70 | 1 777 € |
| Applied Biology Ecology | 23 | 84 | 1 937 € |
| Medical Research | 23 | 77 | 2 185 € |
| Earth sciences Astrophysics | 21 | 72 | 2 020 € |
| Fundamental biology | 18 | 79 | 2 213 € |
| Chemistry | 9 | 64 | 2 230 € |
| Physics | 9 | 45 | 1 976 € |
| Computer Science | 8 | 78 | 2 013 € |
| Engineering | 7 | 84 | 2 138 € |
| Mathematics | 6 | 34 | 1 999 € |

Table 4 shows an asymmetric distribution of the Altmetrics scores according to discipline. The same applies to the average number of readers per article. As can be seen in table 4, the disciplines with the highest Altmetrics scores (after the Multidisciplinary category) are Social sciences and Humanities. These disciplines also have higher than average readership numbers, with the highest number of readers per article averaging 99 in the social sciences. This result is expected insofar as the social sciences and humanities are aimed at a large audience and are more inclined to provoke reactions on the internet, unlike applied disciplines. Mathematics is the discipline with the lowest scores (Altmetrics and number of readers). Other disciplines such as Engineering or Computer science have low Altmetrics scores, but a high average number of readers. Regarding APCs by discipline, the differences are quite small, with a similar distribution to that presented in the previous section (WoS dataset).

Note here that these results may be a consequence of the characteristics of the used database (OpenAPC) and its crossover with the Altmetrics.com and WoS databases.

*Correlation test*

Figure 6 shows the results of Spearman's correlation tests at the journal level (*the crossed out coefficients are not significant at 5%*). As can be seen, the Altmetrics score weakly correlated with the average impact of journals as well as with the average impact of open access publications of these journals. The correlation between APCs and Altmetrics score is low. The correlation between APCs and number of readers is moderate. This suggests that the open access articles that have garnered the most attention on the internet are not necessarily articles for which the authors have paid high APCs. Finally, note that the Altmetrics score is highly correlated with the number of citing posts and accounts.

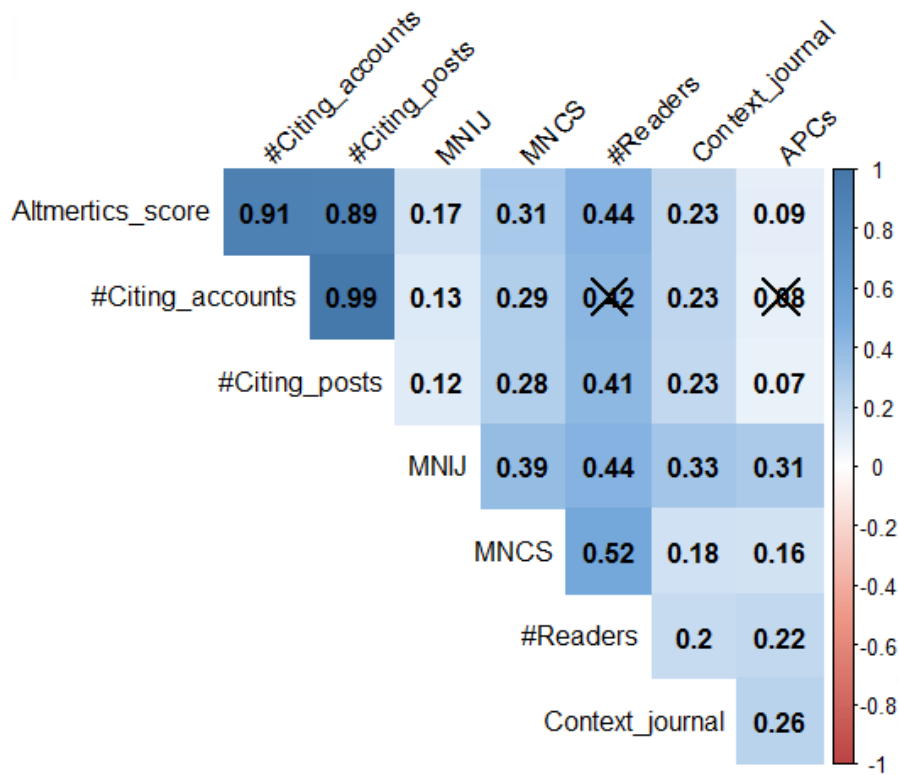

**Figure 6: Spearman correlation matrix at journal level**

*Regression results*

Table 5 shows the results of multiple linear regression. The explained variable is the APCs of 4099 journals. The explanatory and control variables were gradually integrated into models 1, 2, 3 and 4. Several interesting results from table 5 should be highlighted:

**Table 5: Multiple Linear Regression Results**

| | | Dependent variable : log(APCs) of 4099 journals | Model_1 | | Model_2 | | Model_3 | | Model_4 | |
|---|---|---|---|---|---|---|---|---|---|---|
| | | | Coefficient | Pr(>\|z\|) | Coefficient | Pr(>\|z\|) | Coefficient | Pr(>\|z\|) | Coefficient | Pr(>\|z\|) |
| | | Intercept | 7.19*** | <2.22E-16 | 6.89*** | <2.22E-16 | 6.46*** | < 2E-16 | 6.52*** | < 2E-16 |
| Explanatory variables | Journal type | Is_hybrid | 0.46*** | <2.22E-16 | 0.46*** | <2.22E-16 | 0.43*** | < 2E-16 | 0.50*** | < 2E-16 |
| | Impact of journal (log transformed) | MNIJ | - | - | 0.42*** | <2.22E-16 | 0.31*** | < 2E-16 | 0.28*** | < 2E-16 |
| | Altmetrics (log transformed) | Readers count | - | - | - | - | 0.05*** | 1.11E-06 | 0.05*** | 2.60E-06 |
| | | Altmetrics score | - | - | - | - | -0.01* | 6.81E-02 | -0.01* | 9.12E-02 |
| | | Context journal count | - | - | - | - | 0.05*** | < 2E-16 | 0.04*** | 5.66E-09 |
| Control variables | Disciplines (log of the number of papers by discipline) | Humanities | - | - | - | - | - | - | -0.03* | 6.20E-02 |
| | | Mathematics | - | - | - | - | - | - | 0.02 | 5.69E-01 |
| | | Physics | - | - | - | - | - | - | -0.03** | 4.56E-02 |
| | | Applied biology-ecology | - | - | - | - | - | - | 0.01 | 3.38E-01 |
| | | Social sciences | - | - | - | - | - | - | -0.07*** | 3.34E-13 |
| | | Computer sciences | - | - | - | - | - | - | -0.02 | 2.66E-01 |
| | | Earth sciences-astrology | - | - | - | - | - | - | 0.02* | 9.50E-02 |
| | | Engineering | - | - | - | - | - | - | -0.01 | 3.29E-01 |
| | | Medical research | - | - | - | - | - | - | 0.08*** | < 2E-16 |
| | | Chemistry | - | - | - | - | - | - | -0.01 | 5.64E-01 |
| | | Fund biology | - | - | - | - | - | - | 0.03*** | 2.92E-04 |
| Model statistics | | *Ajusted R²* | 0.11 | | 0.20 | | 0.23 | | 0.26 | |
| | | *Dispersion parameter* | 0.24 | | 0.22 | | 0.21 | | 0.20 | |
| | | *Number of Newton-Raphson Iterations* | 2 | | 2 | | 2 | | 2 | |
| | | *AIC* | 5789.5 | | 5338.1 | | 5224.9 | | 5042 | |

*\*\*\* Significant at 1% / \*\* Significant at 5% / \* Significant at 10%.*

- The type of journal plays a huge role in determining APCs. Thus, if the journal is hybrid, the APCs are on average 50% higher than if it is entirely in open access (taking into account all the variables in the model 4). This result may at first glance seem counterintuitive as the business model of fully open access journals is based primarily on the publication fees they charge, which may justify high APCs. Part of the explanation can be found in the fact that some hybrid journals are well-established journals in the market with high impact.
- The link between APCs and the average impact of journals is positive and statistically significant. These results suggest that, overall, the journals take into account their average impact when fixing APCs. In the case of model 4, an increase in impact of 1% is reflected in an increase in APCs of 28%.
- Regarding the societal impact, several lessons can be drawn: the journals do not necessary consider their Altmetrics score when determining APCs. The results even suggest the existence of a slight negative effect of the Altmetrics score on APCs. This would mean that the articles with the highest Altmetrics scores are articles with relatively lower APCs.
- The "Reader count" and "Context journal count" indicators have a positive and significant impact on APCs. This impact remains 6 to 7 times lower than that observed for the academic impact (average impact of journals).

## Conclusion and discussion

Throughout this article, we have analyzed two questions related to APCs. The first one is the relationship between APCs and academic impact. Based on a large sample of 83,752 publications our study empirically verifies the belief that if we pay high costs for publication, impact would necessarily be high. This belief stems from the fact that an author or an institution may think that all publishers who charge a high price for APCs and indexed in international databases like WoS, necessarily have a high academic quality. Our results show that contrary to this belief, paying high costs does not necessarily increase impact of publications. First, large publishers with high impact are not the most expensive in terms of APCs. Second, publishers with highest APCs are not necessarily the bests in terms of impact. Correlation between APCs and impact is moderate.

The second question addressed in this study are the determinants of APCs. More precisely, to what extent the journals take into account their academic and societal impact in determining the APCs. Our results indicate that the type of journal has a strong impact on APCs. Thus, if the journal is hybrid, the APCs are on average 50% higher than if it is entirely in OA. The average impact of the journals also enters, but to a lesser extent, in the determination of APCs. Regarding Altmetrics, the results suggest that they do not have a great impact on APCs. This is particularly the case for the Altmetrics score. The regression results even indicate the opposite. In other words, the OA articles that have garnered the most attention on the internet are articles with low APCs. Another interesting result is that the "number of readers" indicator is more effective and is more correlated with classic impact indicators than the Altmetrics score.

Our results on Altmetrics are consistent with existing studies on the precautions to be taken when using this type of indicator (Sud and Thelwall, 2014; Elmore, 2018; Repiso, Castillo-Esparcia and Torres-Salinas, 2019; Thelwall, 2020). The indicator that best reflects the academic impact is the number of readers (Yang *et al.*, 2021), while the Altmetrics score is weakly correlated with traditional bibliometric indicators (e.g. citation impact). The results of the regression point in the same direction.

Otherwise, in the econometric analysis we have shown that citations score of publications is strongly determined by the impact of journal in which it is published. This result agrees with several studies which show it empirically (Waltman and Traag, 2017). International collaboration also plays an important role in citations score. This result is also consistent with the literature (Larivière *et al.*, 2015).

Another interesting result relates to the impact of hybrid journals versus 100% APCs journals. The regression results indicate that if the journal is hybrid, the NCS is stronger than if it is fully open. This result is consistent with another previous study on the same database (OpenAPC) which showed that journal's impact and hybrid status are the most important factors for the level of APCs (Schönfelder, 2020).

Our results have several implications for public policy and authors choices when it comes to submit their publications. First, the strong interest for OA had an immediate effect on the publishing market. Prices of OA publications have increased exponentially. This increase is disproportionate to the academic impact. The impact of publications for which authors have paid high costs is no better than that of publications with low APCs. Impact may even be lower. We also showed that some publishers are taking advantage of the OA movement to demand high APCs, while their academic impact is very low. Finally, our results suggest that, for authors, APCs should not be used as an indicator for journals selection for submission. For efficient management, institutions should be attentive to journals quality before granting funds for OA publication.

## Limitations

The main limitation of this study is the representativeness of the database. Yet, this should not affect the results concerning the link between APCs, Altmetrics and citation impact, which is our research question and the main object of the study.

Indeed, the citation distribution is homogeneous with that of WoS or other bibliometric databases. Moreover, even if we observe an institutional bias, the database (OpenAPC) includes articles from publishers that represent more than 97% of the market share. As we used averages by publisher, the latter should apply the same prices regardless of the institution. Consequently, this bias should not hide specificities linked to APCs or citation practices. However, we must be careful about the generalization of the results of this article given the representativeness of the database.